\documentclass[manuscript]{aastex}
\usepackage{lscape}
\usepackage{apjfonts}

\shorttitle{A catalog of near-IR unresolved sources}
\shortauthors{Richichi et al.}

\begin{document}


\title{A catalog of near-IR sources found unresolved with milliarcsecond resolution.
\thanks{Based on observations made with ESO telescopes at Paranal Observatory}}


\author{A. Richichi\altaffilmark{1}, O. Fors\altaffilmark{2,3}, F. Cusano\altaffilmark{4}, and M. Moerchen\altaffilmark{5,6}}


\altaffiltext{1}{National Astronomical Research Institute of Thailand, 191 Siriphanich Bldg., Huay Kaew Rd., Suthep, Muang, Chiang Mai  50200, Thailand}
\email{andrea@narit.or.th}
\altaffiltext{2}{Departament Astronomia i Meteorologia and Institut de Ci\`encies del Cosmos (ICC), Universitat de Barcelona (UB/IEEC), Mart\'{\i} i Franqu\'es 1, 08028 Barcelona, Spain}
\altaffiltext{3}{Observatori Fabra, Cam\'{\i} de l'Observatori s/n, 08035, Barcelona, Spain}
\altaffiltext{4}{INAF-Osservatorio Astronomico di Bologna, Via Ranzani 1, 40127 Bologna, Italy}
\altaffiltext{5}{European Southern Observatory, Casilla 19001, Santiago 19, Chile}
\altaffiltext{6}{Leiden Observatory, PO Box 9513, 2300 RA Leiden, The Netherlands}


\begin{abstract}
Calibration is one of the long-standing problems in optical interferometric
measurements, particularly with long baselines which demand stars
with angular sizes on the milliarcsecond scale 
and no detectable companions.
While systems of calibrators have been generally established for
the near-infrared in the bright source regime (K$\la 3$\,mag),
modern large interferometers are sensitive
to significantly fainter magnitudes.
We aim at providing a list of sources found unresolved from
direct observations with high angular resolution and dynamic range,
which can be used to choose interferometric calibrators.
To this purpose, we have used a large number of lunar occultations
recorded with the ISAAC instrument at the VLT to select sources
found to be unresolved and without 
close companions. 
An algorithm has been used to
determine the limiting angular resolution achieved for each source, 
taking into account a noise model built from 
occulted and unocculted portions of the light curves.
We have obtained upper limits on the angular sizes of 556 sources,
with magnitudes ranging from 
 K$_{\rm s} \approx$4 to 10, with a median of 7.2\,mag. The upper limits
 on possible undetected companions (within $\approx 0\farcs5$)
range from K$_{\rm s} \approx$8 to 13, with a median of 11.5\,mag.
One-third of the sources have angular sizes
$\le 1$, and two-thirds $\le 2$ milliarcseconds.
This list of unresolved sources matches well the
capabilities of current large interferometric facilities. 
We also provide available cross-identifications,
magnitudes, spectral types, and other auxiliary information.
A fraction of the sources are found to be potentially variable.
The list covers parts of the
Galactic Bulge and in particular the vicinity of the Galactic
Center, where extinction is very significant and traditional
lists of calibrators are often insufficient.
\end{abstract}





\section{Introduction}
Lunar occultations (LO) can efficiently yield high
angular resolution with short observations using relatively
simple instrumentation. As such, they have been widely employed to
measure  stellar angular diameters and detect small-separation
binary sources, providing the bulk of such information for
several decades.
The  catalog~\citep{CHARM2} listed
several hundreds of LO results. However, the introduction
of large format detectors in the visual and near-IR has
eroded the opportunities at most telescopes
for fast time resolution as required by LO. At the same
time long baseline interferometetry (LBI) has progressively
filled the requests for the same kind of fundamental
observations, without the LO limitations on the choice
of the sources.

LO have recently been re-introduced as a productive
method to obtain high angular resolution
measurements that may not be collected efficiently in
any other way. By using fast integrations on subwindows
of array detectors in the near-IR, LO have been
observed for several hundreds of sources at the
ESO Very Large Telescope (VLT) using a minimal amount of telescope time
\citep[e.g.~][and references therein]{Pleiades}. 
LO targets are selected randomly according to the
lunar apparent orbit, and naturally the majority of
them turn out to be unresolved. Nevertheless, these
unresolved sources are also of interest, in particular
for what concerns interferometry. Indeed, the angular
resolution achieved in these recent LO observations
at the VLT is of order 1 milliarcsecond (mas) and thus
comparable to the typical LBI resolution
on a large interferometer, while the sensitivity is
typically several magnitudes better (K$\approx 12$\,mag).

Therefore, sources which are found to be unresolved
by LO are well suited  to be used as calibrators by 
near-IR LBI.
Considering that in our experience about 10\% of the randomly-selected
LO sources turn out to be resolved or binary, and that
so-called "bad" LBI calibrators are a concern~\citep{hummel}, 
such a selection of clean calibrators is indeed desirable.
In this work, we list 556 sources found to be
unresolved from our LO observations at the VLT, and
thus highly homogeneous in terms of data and
analysis. For each, we provide an upper limit on
the angular size as well as a sensitivity limit on the
possible presence of undetected companions.
The sources are mainly located in the Galactic Bulge,
with several in the vicinity of the Galactic Center
where visual extinction can be extreme and therefore
it might be more difficult to identify suitable
LBI calibrators.

\section{Observations and data analysis}\label{data}
The observations which form the basis of this work
were obtained  using the ISAAC instrument~\citep{ISAAC} 
at the ESO VLT. The instrument
was installed at the 8.2-m UT1 Antu telescope until May 2009,
and thereafter at the twin UT3 Melipal telescope.
The majority of the observations were carried out in 
service mode following a {\it filler} strategy to utilize
small amounts of time available between other programs
or due to unfavorable atmospheric conditions \citep[e.g.,~][]{richichi2010}. 
A significant amount of observations
was also carried out in visitor mode \citep[e.g.,~][]{richichi2011}. 

Each observation consisted of several thousands of frames
in a small sub-window with short sampling (also equal to integration) times.
Typical values were 7000 frames of 32x32-pixel ($4\farcs7 \times 4\farcs7$) at 3.2\,ms,
although different combinations exist across our data set.
A broad-band K$_{\rm s}$ filter was generally
employed, except in the case of very bright sources
for which a narrow-band filter centered at 2.07~$\mu$m was used. 
The events were generally disappearances, although a
number of reappearances were also observed.
As expected from the nature of the data set, the observing
conditions covered a very wide range, with median values of
seeing, airmass, lunar phase being $0\farcs88$, 1.30 and 63\%,
respectively. The final quality of the light curves
is only marginally affected by atmospheric conditions, however
other important factors
are the image quality (LO at the VLT 
are observed with the primary mirror active optics off)
and the intensity of the lunar background. 
These can vary significantly from
one LO event to another and may be responsible for a large
scatter in signal-to-noise ratio (SNR) among light curves of
sources with similar magnitude.

The data cubes were converted to light curves using
a mask extraction, and these in turn were analyzed using both
a model-dependent~\citep{richichi96} 
and a model-independent method \citep[CAL,~][]{CAL}.
Details on the instrumentation and the method
can be found in~\citet{richichi2011} 
and references therein. Large samples of LO data inevitably
include occasional cases of light curves with slightly distorted
diffraction patterns, most likely as a result of irregularities
in the lunar limb. This appears to be statistically more frequent
in observations with a large telescope which samples a wider
section of the limb and which also provides an increased sensitivity.
\citet{richichi2012} discussed the nature and
the effect of such irregularities, and suggested correction algorithms
that we employed for some light curves in our sample.

\begin{figure}
\includegraphics[angle=-90.0, width=15cm]{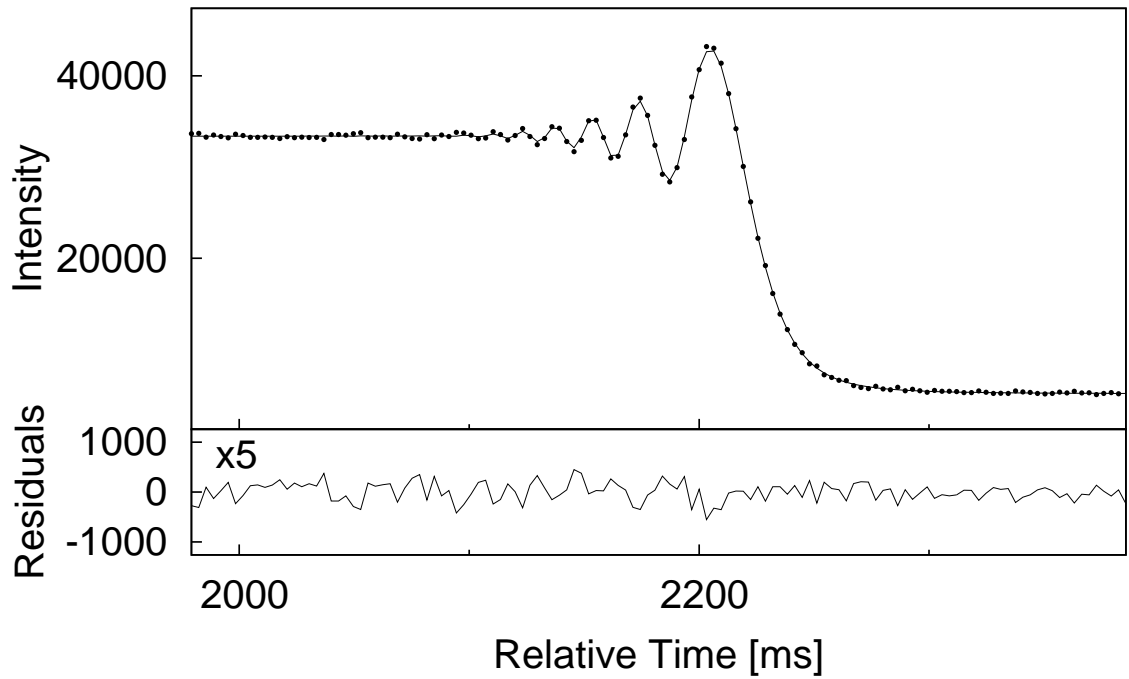}
\caption{
The top panel shows the occultation light curve (dots) 
and the point-source fit (solid line)
for 2MASS 17383557-2223159, a K=6.0\,mag source without known
cross-identifications. The SNR of the fit is 148 and the
limiting angular resolution is $\phi_{\rm u}=0.50$\,mas.
The lower panel shows the fit residuals, enlarged 
for clarity.
}
\label{fig_fit1}
\end{figure}

\begin{figure}
\includegraphics[angle=-90.0, width=15cm]{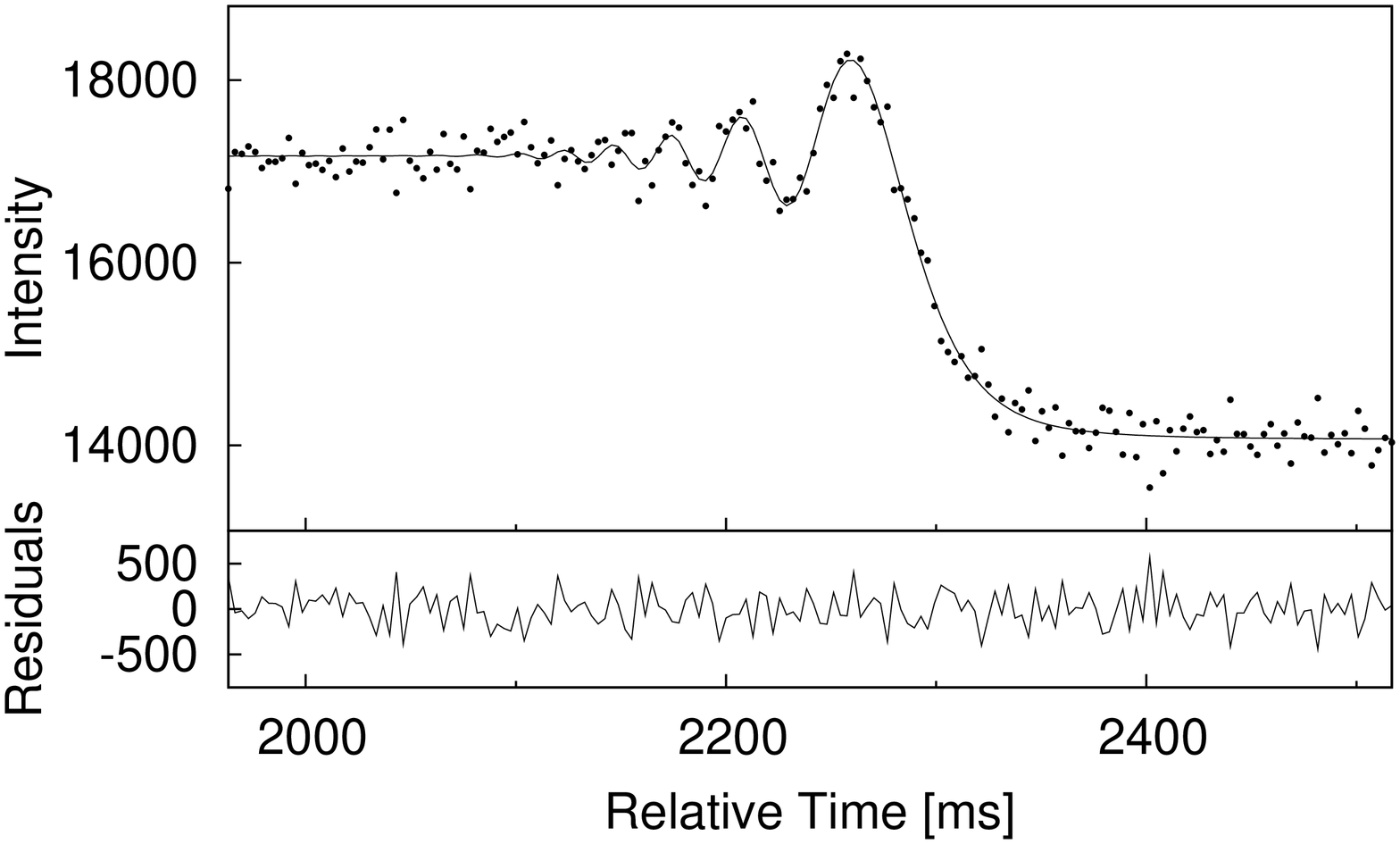}
\caption{Same as Fig.~\ref{fig_fit1}, for 2MASS 06251004+2302141,
a K=8.2\,mag source coincident with TYC~1879-97-1.
The SNR of the fit is 16 and the
limiting angular resolution is $\phi_{\rm u}=3.05$\,mas.
}
\label{fig_fit2}
\end{figure}

An important aspect of our data reduction was the systematic
assessment of the limiting angular resolution achieved for
each of the light curves. This was based on an algorithm
which evaluates the quality of the fit
over a range of angular diameters, taking into account
also factors such as the finite number of data points and
their intrinsic numerical accuracy. Details are provided
in~\citet{richichi96}. This quantity, which
we denote with $\phi_{\rm u}$, can be considered equivalent
to an upper limit on the angular size of the source.
In Fig.~\ref{fig_fit1} we show an example of the data
and best fit by a point-source for the case of a
high SNR light curve, while Fig.~\ref{fig_fit2} provides
a similar exmaple for a low SNR case.

For each light curve, we combine the SNR
and the K magnitude to derive 
the detection limit K$_{\rm lim}$ for a hypothetical companion.
In a LO event the noise level might
not be symmetric around the time of occultation. Also, depending
on the circumstances and geometry, detections of companions
below the SNR=1 level can be possible. Therefore, the K$_{\rm lim}$
quantity is not a strict limit but rather an indication.
We note that
only restricted portions of the light curves,
corresponding to angular extensions of $\approx 0\farcs5$, were analyzed. 
For our scopes, the definition of unresolved source is only
meaningful within this restricted range.
Companions
further away than this limit have been noticed in some
cases or might have been present without being detected, however we
do not concern ourselves with them.




\section{Source selection}\label{sample}
A total of 915 successful LO events
were logged in our database from March 2006 until June 2012.
A significant fraction of these observations yielded positive
results in areas of binary stars, stellar diameters, and sources
with extended circumstellar emission are described in other
papers already published or in preparation. 
Here, we concern ourselves only
with the sources effectively found to be unresolved.
Several tens of sources already found to be resolved or binary
during our observations at the VLT
have already been published \citep[e.g.~][and references therein]{Pleiades}.

The first step in their selection was
the rejection of all light curves showing
binary or multiple stars, within the angular range mentioned 
in Sect.~\ref{data}. This included borderline
cases in which the CAL algorithm revealed a secondary peak
in the brightness profile, generally close to the noise level, but for
which no significant improvement in the quality of the fit could be
reached with a double-star model.
We then rejected all stars with a resolved angular diameter or an otherwise extended apperance,
including those for which the $\phi_{\rm u}$ algorithm yielded
an indication of being resolved even though no diameter could be fitted.
We also rejected light curves with significant distortions of
the fringe pattern due to presumed limb irregularities, e.g., exceeding
by several times the noise level over several fringes. 
However, moderate distortions were generally satisfactorily
treated with the algorithms of~\citet{richichi2012} and
the corresponding light curves were retained.
Finally, a minimum SNR of 10 was imposed.

As a result, a total of 556 sources were selected and are listed
in Table~\ref{Table1}, available online.
We used
the 2MASS Catalogue for our predictions, and
further
identifications are extracted from the {\it Simbad} database.
We also provide our own ID number that combines
the ESO Period number with the entry in our database.
This ID also marks with an asterisk sources observed with a narrow band filter.
Further we list coordinates,
 BVJHK photometry, spectrum, number of references and
source type, as available from {\it Simbad}.
The next fields in the Table are the SNR, K$_{\rm lim}$ and $\phi_{\rm u}$
as described in Sect.~2 and computed by us. 
Finally, we flag sources known or likely to be variable.
By the flag S we denote sources which belong to categories such
as Miras, emission-line stars, RR Lyr or
other kind of variables. By the flag F we denote stars for which
our measured counts differed from the counts expected on the
basis of the 2MASS K$_{\rm s}$ value by more than 0.6\,mag
(see also Fig.~\ref{fig_counts}). 

\section{Discussion}\label{discussion}
\begin{figure}
\plotone{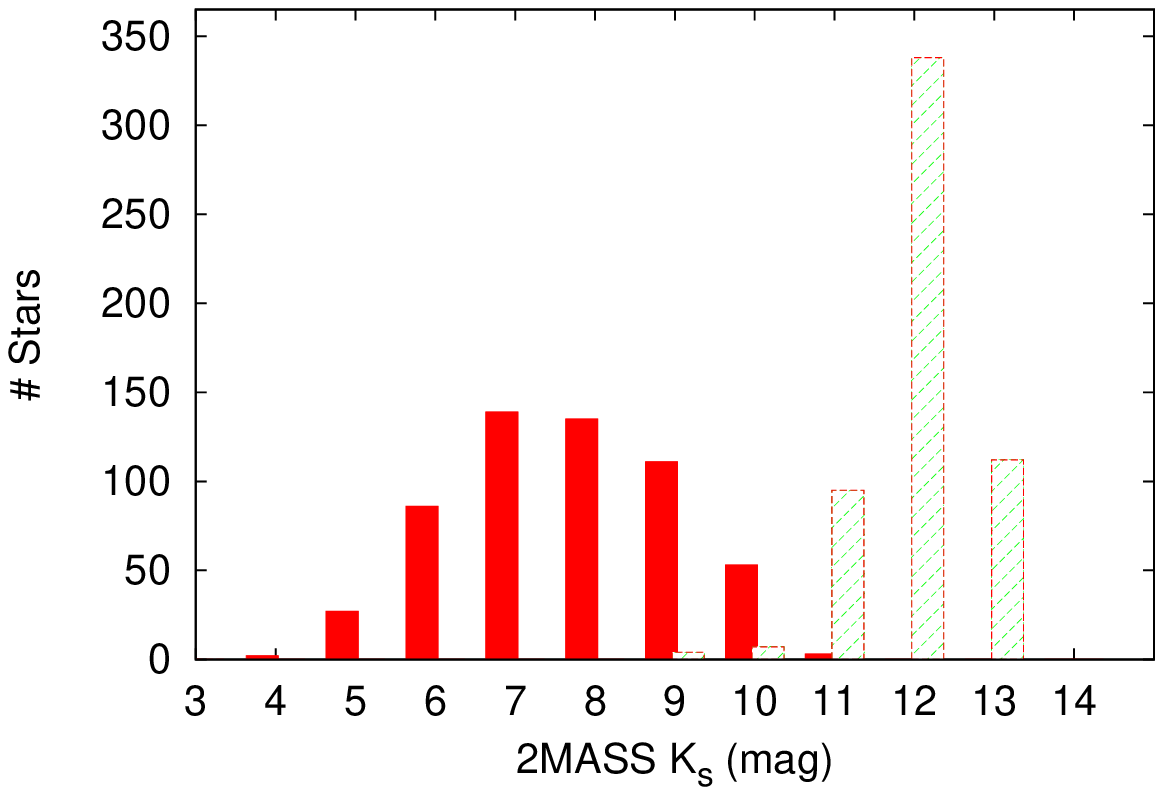}
\caption{Distribution of 2MASS K$_{\rm s}$ magnitudes (red, filled)
and limiting magnitudes K$_{\rm lim}$ (green, striped)
for the sources in our list.}
\label{fig_km}
\end{figure}

The distribution in magnitude for the sources in our list
is shown in Fig.~\ref{fig_km}. We used the nominal K$_{\rm s}$
values from 2MASS, although our measurements have
shown at times significantly different values as discussed later.
They range from K$_{\rm s}$=3.7 to 10.2, with a median of 7.2\,mag.
The figure also shows the distribution
of the limiting magnitudes K$_{\rm lim}$ as introduced in Sect~\ref{data}.
The values range from K$_{\rm s}$=8.2 to 12.8, with a median of 11.5\,mag.

\begin{figure}
\plotone{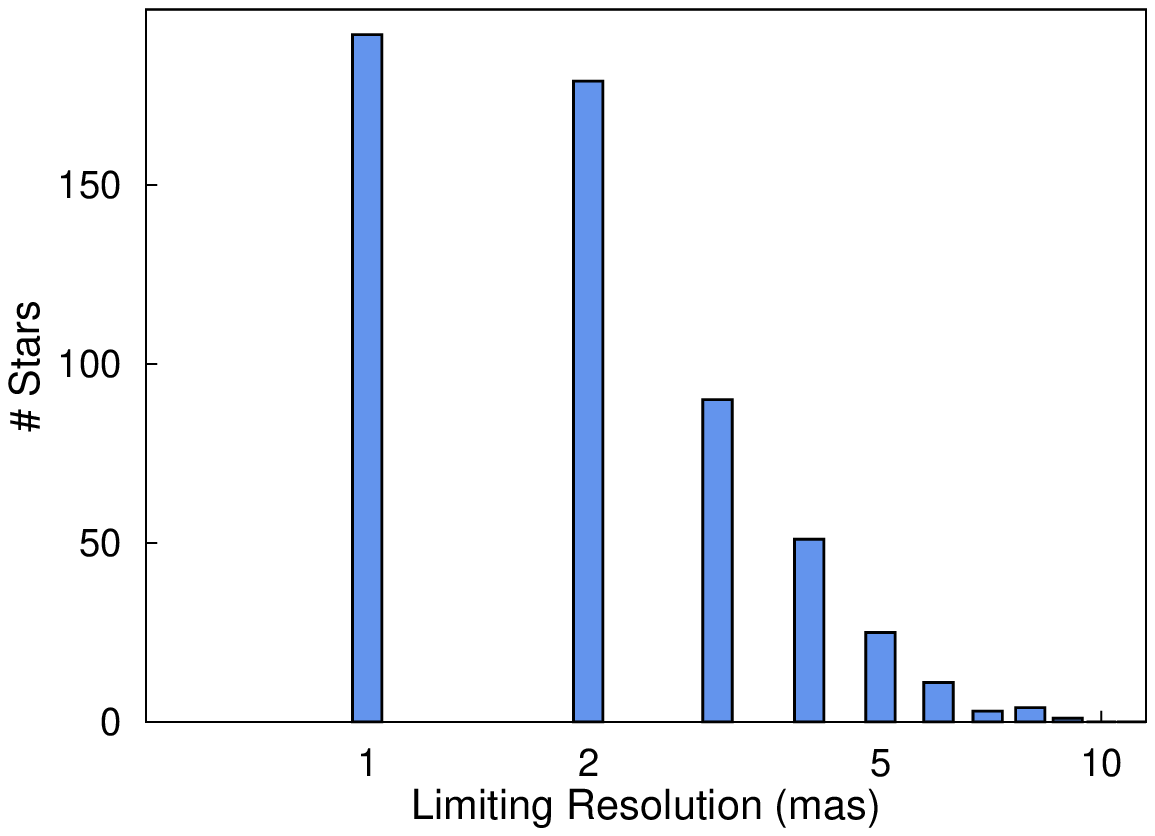}
\caption{Distribution of limiting angular resolution, $\phi_{\rm u}$,
for the sources in our list.}
\label{fig_limres}
\end{figure}

\begin{figure}
\plotone{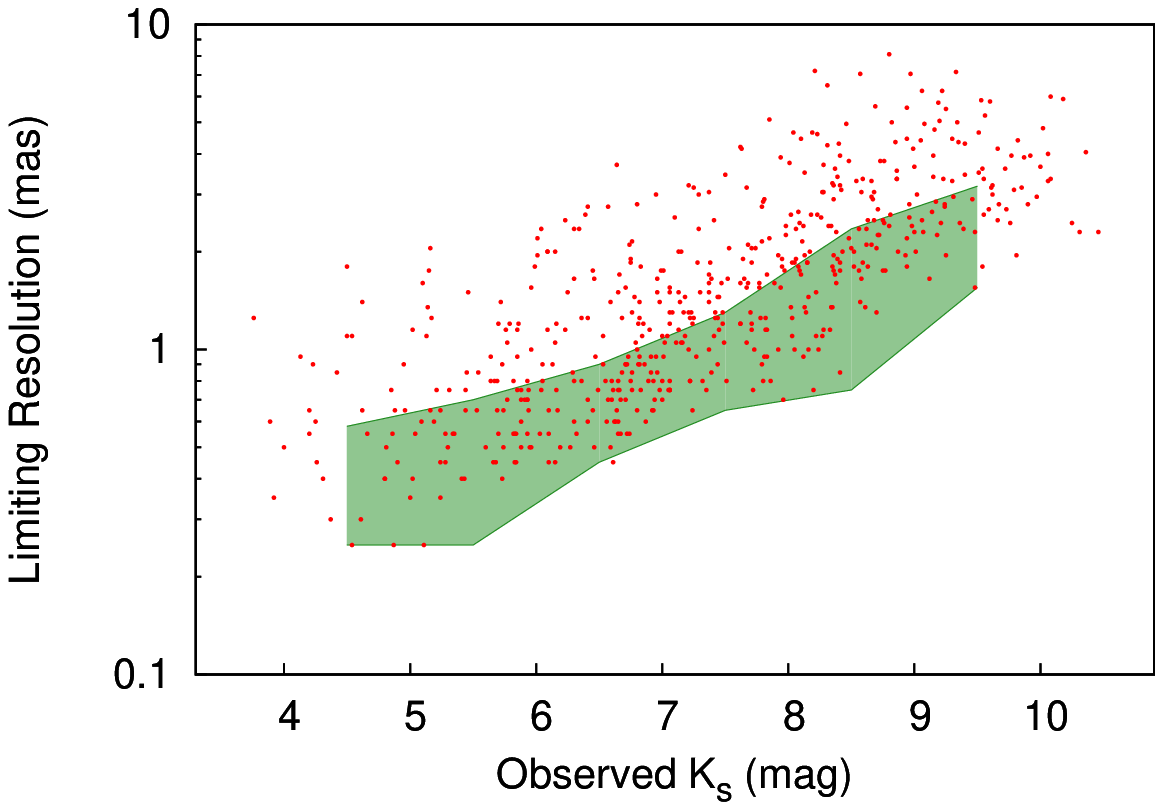}
\caption{Distribution of limiting angular resolution $\phi_{\rm u}$
for the sources in our list, as a function of the observed
magnitude. The shaded area is delimited at the bottom and at the top by the minimum and
the median, respectively, of the $\phi_{\rm u}$ values over one-magnitude bins.\label{fig_limres2}}
\end{figure}

\begin{figure}
\plotone{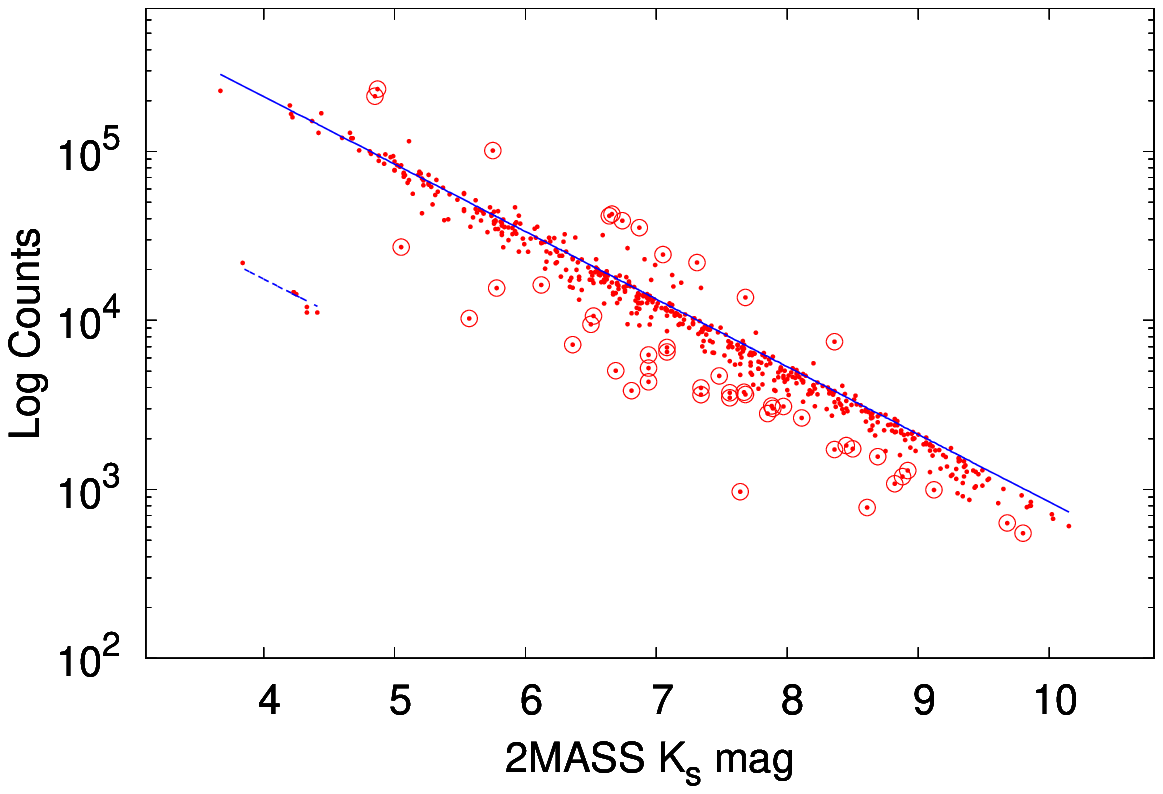}
\caption{Expected (lines) and measured (dots) flux values
for the 2MASS K magnitudes of the sources in our list. The expected
values are based on the ISAAC exposure time calculator (solid line for
broad-band K$_{\rm s}$, and dashed line for narrow-band 2.07\,$\mu$m filters).
The circled dots denote sources which we flag as variable, see text.}
\label{fig_counts}
\end{figure}

The distribution of the upper limits on the angular sizes $\phi_{\rm u}$
is shown in Fig.~\ref{fig_limres}. The values range from
0.25 to 8.1\,mas, with a median of 1.4\,mas. A total of 192 sources,
or about one third of the sample, have a limiting angular resolution
of 1.0\,mas. The $\phi_{\rm u} \la 2.0$\,mas limit is satisfied for 371 sources.
Fig.~\ref{fig_limres2} shows another view of the $\phi_{\rm u}$ distribution, 
this time as a function of the K$_{\rm s}$ magnitude. For this, we converted the
measured counts into an observed K$_{\rm s}$ magnitude, using the
ISAAC exposure time calculator (see also Fig.~\ref{fig_counts}).
The shaded area in Fig.~\ref{fig_limres2} covers the range between
the minimum and the median $\phi_{\rm u}$ values 
over one-magnitude bins. The scatter in $\phi_{\rm u}$
is a reflection of different SNR encountered for a same
source brightness depending on observing conditions, as mentioned in Sect.~\ref{data}.
However, when just the best 50\% of the sources are considered in each
magnitude bin, the LO performance appears to be quite consistent and
shows that the 1\,mas limit is attained up to K$\approx 7$\,mag.

It is interesting to compare our upper limits $\phi_{\rm u}$
with the predicted angular diameters.
Several empirical relations have been derived, to estimate a star's diameter
from its colors. A recent one by \citet{kerv04} is based on V and K photometry and
can be applied to dwarf stars in the spectral range A0 to M2, and sub-dwarf stars 
between A0 and K0. Unfortunately, we face the complication that several of our
stars suffer from considerable reddening, have little spectral information,
and in some cases appear to be variable -
as explained in the following.
Nevertheless, we have endeavoured to determine the estimated diameter for
a subset of 90 stars within the above constraints of photometry and spectral
class, without taking into consideration possible interstellar reddening.
The result is that in all cases the $\phi_{\rm u}$ is larger than, or at the most
approximately similar to, the estimated diameter. Only in one case we found
some discrepancy (our upper limit being about 25\% larger
than the estimated diameter). This is IRAS~14434-2055, a source with over
4 magnitudes of V-K color, which is likely not intrinsically photosperic.
We conclude that, as far as it is possible to check, our upper limits
are well consistent with the expected stellar diameters.

\begin{figure}
\begin{center}
\includegraphics[angle=-90.0, width=15cm]{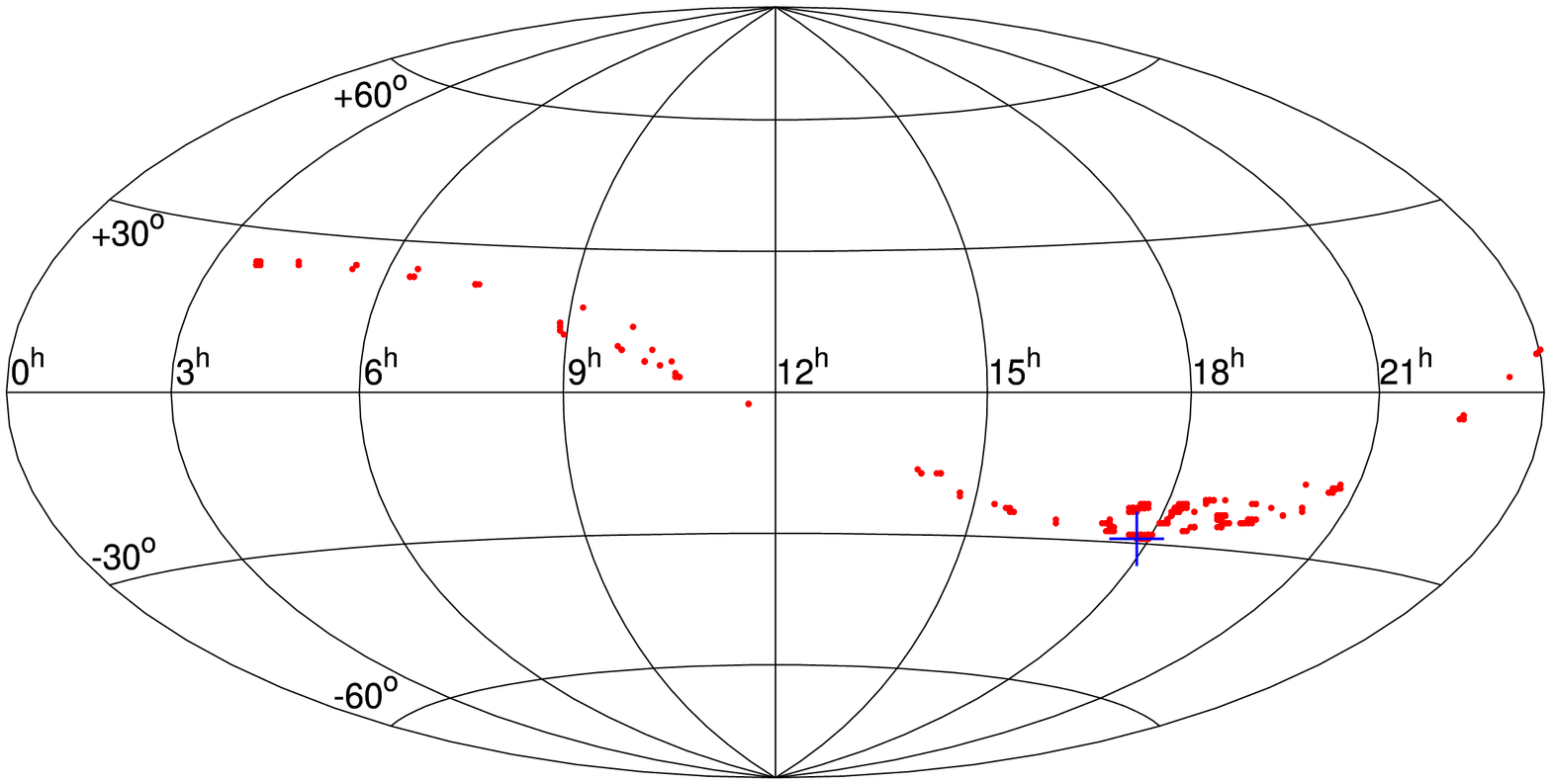}
\caption{Sky distribution of the sources in our list.
The Galactic Center is marked with a cross.}
\label{fig_sky}
\end{center}
\end{figure}

\begin{figure}
\plotone{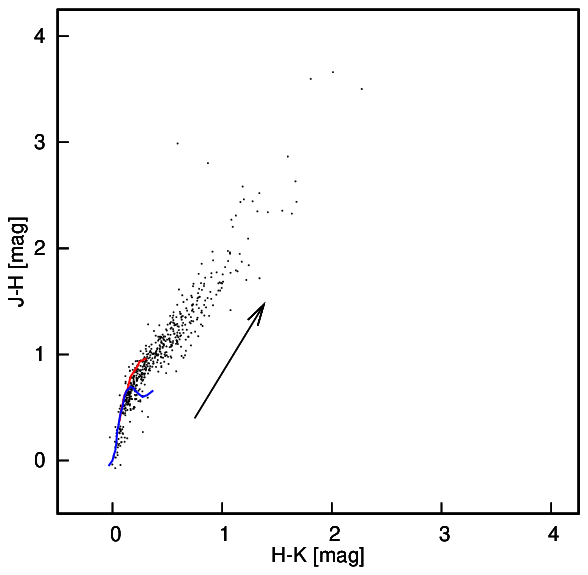}
\caption{2MASS color-color diagram for the sources in our list.
The lines are the loci of the unreddened giant (red) and
dwarf (blue) stars, respectively, according to~\citet{bessellbrett}.
The arrow is the extinction vector for A$_{\rm V}$=10\,mag, according to
~\citet{riekeleb}.}
\label{fig_colors}
\end{figure}

The majority of the sources have no known counterpart,
with just under 50\% having any identification other
than the 2MASS name. About 30\% have at least
one bibliographical entry, and only 13\% have at least two.
Spectral types are known for only 15\% of the sources,
and V magnitudes for only one out of three.
This is a reflection of many of the sources being in
the Galactic Bulge, see Fig.~\ref{fig_sky},
where interstellar extinction
is significant or even extreme.
As a result, the sources in our list are also mainly red or very
red in color, as shown in Fig.~\ref{fig_colors}. While generally
consistent with a giant sequence, the near-IR colors show that
many of the sources are affected by 10 or even 20 magnitudes
of visual extinction. 

In fact, we note  that 10 sources are within 1$\degr$ of the
Galactic Center, and 45 are within 5$\degr$. This highlights a
potential use of our list, namely to provide reference 
stars for LBI near-IR observations in a very
interesting area of the sky,
for which calibrators selected
 by other methods such
as the modelling of the spectral energy distribution~\citep{Cohen} 
are very hard to find.

Fig.~\ref{fig_counts} shows the average unocculted counts within
our extraction masks for all sources, as a function of the nominal
2MASS K$_{\rm s}$ magnitudes. It can be observed that in general
the counts follow very closely the theoretical ISAAC performance.
However, in a significant number of cases deviations are observed
both in the sense of the observations showing brighter or fainter
magnitudes than 2MASS. Small variations can be attributed to
airmass (which we did not account for), atmospheric conditions and to
the mask extraction process. However, larger deviations are
likely indicative of intrinsic variability. In Fig.~\ref{fig_counts}
we have used 0.6\,mag as a threshold, leading to about 9\% of the
sources appearing to have varied their brightness since the
2MASS measurement. 

The problem of finding suitable calibrator stars in LBI
has been long-standing. \cite{Cohen} 
produced an extensive compilation of predicted angular
diameters, based on modelling
of the spectrophotometric properties of bright non-variable stars.
This has often been used as a source of calibrators, and
variations more suitable for the specific application
of near-IR LBI have also been produced 
\cite[e.g.,~][]{merand}. More recently,
as extensive databases of observations
have become available from facilities such as the
Palomar Testbed Interferometer and the ESO Very Large
Telescope Interferometer, attempts have been made
to derive self-consistent systems of calibrators with
measured, as opposed to predicted, diameters
\citep{vanbelle,richichi2009}.

While the choice of suitable calibrators is largely eased
by the above mentioned works, a few limitations are
still encountered: for example, the density of calibrators 
is less than satisfactory in some areas of the sky, and 
they are usually limited to rather bright stars,
typically K$< 3$\,mag. Tools for selecting fainter calibrators
are available \cite{bonneau};
however they are again based on indirect methods and
they may require 
interpolations for missing photometry. The quoted photometric 
accuracy of 0.1\,mag is generally acceptable to estimate
angular diameters, and large relative errors are not 
a problem in the faint magnitude (small size) regime.
However, such accuracy does not allow us
to ascertain in advance, for example, whether a specific
source has a faint companion. On the other hand, experience from 
our LO observations has shown that, with a resolution of $\approx 1$\,mas and
a dynamic range of $\Delta$K$\approx 5$\,mag, about 10\% of
the field stars appear to have companions.

Interferometric observations are generally quite time-consuming
both in acquisition and analysis of the data, and it is often
only after the observing run has been concluded that problems
connected with a calibrator, such as a previously unknown
companion, can be recognized. 
In this sense, our list of unresolved sources can prove
useful. Admittedly, we  provide only upper
limits rather than angular diameters, however for baselines
which are not too long this can be sufficient. For example,
on a 50\,m baseline in K band, the difference in squared visibility
between a 1\,mas and a point-like source is 3\%, comparable
to the typical accuracy of an interferometric measurement.
Of course, LBI observations repeated on different baselines
can detect deviations in the point-like nature of a source with
better accuracy. 
The case of an unknown faint companion, however, might be harder to recognize.

\section{Conclusions}
We have used a database of about 900 near-IR lunar occultation light curves
obtained with the ISAAC instrument at the VLT, to identify a set
of sources that can be classified as unresolved.
This set comprises 556 stars spanning a range of magnitudes from
K$\approx$4 to 10, which have been analyzed in a
highly homogeneous context in terms of instrument performance 
and data analysis. 
We are able to set upper limits on their
angular sizes, typically of 1 to 2 milliarcseconds, and on
the presence of nearby companions,  with an average dynamic
range of 4.5\,mag, reaching a typical limiting
magnitude K$\approx$11 to 12.
The sources are distributed along the Zodiacal belt, and especially
clustered in the Galactic Bulge. They exhibit mostly red or
very red colors due to interstellar extinction. A significant number are
in close proximity to the Galactic Center. As a result, only a small fraction
of the sources in our list are well studied, with the majority lacking
extensive classification and bibliographical references.

We identify a key application of this list as a database of
calibrators for long-baseline interferometry, especially in the
faint source regime of modern large facilities and for targets
in the general direction of the Galactic Center where other
sources of reliable calibrators are scarce.

\acknowledgments{
The observations which form the basis of this work would
not have been possible without the support and dedication
of the ESO staff, both in Garching and in Chile, whom we
thank sincerely.
OF acknowledges financial support from MICINN through 
a {\it Juan de la Cierva} fellowship and from 
\emph{MCYT-SEPCYT Plan Nacional I+D+I AYA\#2008-01225}. 
MM acknowledges co-funding under the Marie Curie Actions 
of the European Commission (FP7-COFUND) through the ESO Fellowship program.
This research made use of the Simbad database,
operated at the CDS, Strasbourg, France, and
of data products from the Two Micron All Sky Survey (2MASS), 
which is a joint project of the University of Massachusetts 
and the Infrared Processing and Analysis Center/California Institute 
of Technology, funded by the National Aeronautics and 
Space Administration and the National Science Foundation.
}




\clearpage

\clearpage
\begin{landscape}
\begin{deluxetable}{ccccccccccccccccc}
\tabletypesize{\scriptsize}
\tablecaption{List of sources found unresolved by the lunar occultation technique\label{Table1}}
\tablewidth{0pt}
\tablehead{
\colhead{Object}&\colhead{Our ID}&\colhead{Simbad ID}&
\colhead{${\alpha}$(J2000)}&\colhead{${\delta}$(J2000)}&
\colhead{B}&\colhead{V}&\colhead{J}&\colhead{H}&\colhead{K}&
\colhead{Sp}&\colhead{\#Ref}&\colhead{Type}&\colhead{SNR}&
\colhead{K$_{\rm lim}$}&\colhead{$\phi_{\rm u}$}&\colhead{Var}
}
\startdata
17383141-2812445	 &	 P76/gc097         & \object{DENIS-PJ173831.3-281244}         &   17:38:31.4  & -28:12:44.5  &         &          &   8.88    &   7.47  &  6.78  &            & 1    & IR                                         & 70.8   &  11.4 & 1.50   &      \\
17385533-2836105	 &	 P76/gc137         & \object{2MASSJ17385533-2836105}          &   17:38:55.3  & -28:36:10.5  &         &          &   9.96    &   8.33  &  7.34  &            & 1    & IR                                         & 44.8   &  11.5 & 1.80   &      \\
17385407-2833227	 &	 P76/gc147         & \object{DENIS-PJ173854.0-283322}         &   17:38:54.1  & -28:33:22.7  &         &          &   9.80    &   8.03  &  7.13  &            & 1    & IR                                         & 47.5   &  11.3 & 1.20   &      \\
17394022-2820124	 &	 P76/gc199         & \object{2MASSJ17394022-2820124}          &   17:39:40.2  & -28:20:12.4  &         &          &   9.81    &   8.12  &  7.12  &            & 1    & IR                                         & 84.3   &  11.9 & 3.70   &      \\
17394602-2822089	 &	 P76/gc207         & \object{2MASSJ17394602-2822089}          &   17:39:46.0  & -28:22:08.9  &         &          &   9.72    &   7.87  &  7.05  &            & 1    & IR                                         & 92.8   &  12.0 & 2.35   & F    \\
17405112-2817130	 &	 P76/gc260         & \object{2MASSJ17405112-2817130}          &   17:40:51.1  & -28:17:13.0  & 19.10   &          &   7.07    &   5.51  &  4.87  &            & 1    & IR                                         & 181.7  &  10.5 & 0.60   & F    \\
17411415-2818051	 &	 P76/gc284         & \object{2MASSJ17411415-2818051}          &   17:41:14.2  & -28:18:05.1  & 18.50   &          &   8.88    &   7.49  &  6.87  &            & 1    & IR                                         & 100.4  &  11.9 & 0.75   & F    \\
17413435-2829225	 &	 P76/gc352         & \object{ISOGAL-PJ174134.6-282922}        &   17:41:34.3  & -28:29:22.5  &         &          &   6.96    &   5.52  &  4.85  &            & 5    & Star,IR                                    & 241.7  &  10.8 & 0.50   & F    \\
17425497-2836593	 &	 P76/gc426         & \object{2MASSJ17425497-2836593}          &   17:42:55.0  & -28:36:59.3  &         &          &   10.49   &   7.90  &  6.72  &            & 1    & IR                                         & 92.8   &  11.6 & 0.80   &      \\
17430901-2837039	 &	 P76/gc437         & \object{2MASSJ17430901-2837039}          &   17:43:09.0  & -28:37:03.9  &         &          &   10.59   &   8.16  &  6.99  &            & 1    & IR                                         & 85.7   &  11.8 & 0.75   &      \\
\enddata
\tablecomments{Table~\ref{Table1} is published in its entirety in the 
electronic edition of the {\it Astrophysical Journal Supplement Series}. A portion is 
shown here for guidance regarding its form and content.}
\end{deluxetable}
\clearpage
\end{landscape}


\clearpage






\end{document}